\documentclass[prd,aps,preprintnumbers,amsmath,amssymb,nofootinbib,eqsecnum, preprintnumbers]{revtex4}
 \pdfoutput=1
\usepackage{graphicx}
\usepackage{epsf}
\usepackage{amsmath}
\usepackage{epstopdf}
\usepackage{bm}
\usepackage{color}
\usepackage{graphicx}
\usepackage{subcaption}
\usepackage{tabularx}
\usepackage{float}
\usepackage{footnote}
\usepackage{graphicx}
\usepackage{hyperref}
\usepackage{amssymb,epsf}
\usepackage{latexsym}
\usepackage{epstopdf}
\usepackage{epsfig}
\begin{document}

\title{Information Theoretical Approach to Detecting Quantum Gravitational Corrections}

\author{Behnam Pourhassan$^{1,2}$
\footnote{email address: b.pourhassan@du.ac.ir},
Xiaoping Shi$^3$,
\footnote{email address: xiaoping.shi@ubc.ca},
Salman Sajad Wani$^{4}$
\footnote{email address: sawa54992@hbku.edu.qa},
Saif-Al-Khawari$^{4}$
\footnote{email address: smalkuwari@hbku.edu.qa },
Farideh Kazemian$^{1}$
\footnote{email address: fkazemian.144@gmail.com},
\.{I}zzet Sakall{\i}$^{5}$
\footnote{email address: izzet.sakalli@emu.edu.tr},
Naveed Ahmad Shah$^{6,7}$
\footnote{email address: naveed179755@st.jmi.ac.in},
Mir Faizal$^{7,8,9,10}$
\footnote{email address: mirfaizalmir@googlemail.com}
}
\affiliation{$^1$School of Physics, Damghan University, Damghan 3671645667, Iran.}
\affiliation{$^2$ Center for Theoretical Physics, Khazar University, 41 Mehseti Street, Baku, AZ1096, Azerbaijan.}
\affiliation{$^3$Department of Computer Science, Mathematics, Physics and Statistics, University of British Columbia, Kelowna, BC V1V 1V7, Canada.}
\affiliation{$^4$ Qatar Center for Quantum Computing, Hamad Bin Khalifa University, Doha, Qatar.}
\affiliation{$^5$Physics Department, Eastern Mediterranean University, Famagusta 99628, North Cyprus via Mersin 10, Turkey.}
\affiliation{$^6$Department of Physics, Aligarh Muslim University, Aligarh- 202002, India.}
\affiliation{$^7$Canadian Quantum Research Center 204-3002 32 Ave Vernon, BC V1T 2L7 Canada.}

\affiliation{$^8$Irving K. Barber School of Arts and Sciences, University of British Columbia Okanagan, Kelowna, BC V1V 1V7, Canada.}

\affiliation{$^9$ Department of Mathematical Sciences, Durham University,
Upper Mountjoy,  Stockton Road, Durham DH1 3LE, UK. } \affiliation{$^{10}$ Faculty of Sciences, Hasselt University,  Agoralaan Gebouw D, 3590 Diepenbeek, Belgium.}

\begin{abstract}
In this paper, we investigate the scales at which quantum gravitational corrections can be detected in a black hole using information theory. This is done by calculating the Kullback-Leibler divergence for the probability distributions obtained from the Parikh-Wilczek formalism. We observe that as  quantum gravitational corrections increase with decrease in scale, the  increase the Kullback-Leibler divergence  between the original and quantum gravitational corrected probability distributions will also increase. To understand the impact of such quantum gravitational corrections we use Fisher information. We observe that it again increases as we decrease the scale.   We obtain these results for higher-dimensional black holes and observe this behavior for Kullback-Leibler divergence and Fisher information also depending on the dimensions of the black hole. Furthermore, we observe that the Fisher information is bounded and  approaches a fixed value. Thus, information about the nature of quantum gravitational corrections itself is intrinsically restricted by quantum gravity. Thus, this work establishes an intrinsic epistemic boundary within quantum gravity. 
\end{abstract}

\maketitle

\section{Introduction}
\label{intro}
Various proposals for quantizing gravity lead to different theoretical modifications of low-energy quantum phenomena \cite{le1, le2, le4, le5, le6, le7}. Thus, it is important to detect effects produced by quantum gravity, and various tests have been proposed for such quantum gravitational effects \cite{qge12, qge14}. Among them, it is speculated that black hole physics can be used to test quantum gravitational effects \cite{qge15, qge16}. Although several tests for quantum gravity have been proposed, the scale at which quantum gravity effects are likely to be observed has not been rigorously discussed. 
To properly classify and quantify the dependence of quantum gravitational effects on the scale, we analyze the probability distribution of particles emitted from a black hole in the Parikh-Wilczek formalism \cite{pw, pw1}. As the Parikh-Wilczek formalism considers the back reaction of particles emitted during black hole evaporation, it is sensitive to changes in the geometry due to quantum gravitational effects. This feature of the formalism thus allows the determination of the change in the probability distribution of the emitting particles due to these effects. To quantify the scale dependence of such quantum gravitational effects, we use information-theoretical techniques. We start by using the Kullback-Leibler divergence \cite{kl1, kl2}, which measures the difference between two statistical probability distributions. While not a metric due to its asymmetry,  Kullback-Leibler divergence effectively quantifies the deviation between two distributions. Thus, a higher Kullback-Leibler divergence would mean that the corrected probability distribution would have a higher deviation from the original probability distribution. This, in turn, would make it easier to experimentally detect the effects of such corrections. Accordingly, the Kullback-Leibler divergence can clearly quantify and classify the effects of quantum gravitational corrections. 

We observe that Kullback-Leibler divergence   increases as the mass of the black hole decreases. We claim that this is due to quantum gravitational effects. To quantify this claim, we use Fisher information, which effectively quantifies information about a parameter that can be obtained from a given distribution \cite{fish1, fish2}. We thus directly use Fisher information to analyze the information we obtain about quantum gravitational corrections. We demonstrate that Fisher information about quantum gravity also  increases as the mass decreases, and is bounded by a certain critical value.  Moreover, we find that  the Kullback-Leibler divergence and Fisher information depends on dimensions.    Our approach allows us to thoroughly evaluate the influence of quantum gravity on black holes across various scales and dimensions. This can be directly related to the possibility of detecting quantum gravitational corrections. We also note that the scale dependence of quantum gravitational effects in higher dimensions has already been discussed \cite{k1, k2, b1, b2}. This observation has motivated the study of higher-dimensional Schwarzschild black holes  \cite{ed1, ed2, ed3, ed4}. Here, we analyze the effects of quantum gravitational corrections using such higher-dimensional Schwarzschild black holes. This allows us to explicitly investigate the effects of dimensions on quantum gravitational corrections, which can be determined by using an effective quantum corrected metric and by introducing the concept of a novel quantum mass \cite{j6}. This quantum mass becomes important in analyzing changes in the probability distribution of emitted particles in the Parikh-Wilczek formalism \cite{pw, pw1}.

The phenomenological consequences of quantum gravitational corrections to black hole thermodynamics have been used to propose tests of quantum gravitational effects \cite{qge17, qge18, qge19, qge10}, noting that the standard black hole thermodynamics is a semi-classical theory, which is obtained using quantum field theory in curved spacetime 
  \cite{1a00, 1b, 1c, 1a0}. In this approach, the thermodynamic properties of black holes are obtained by neglecting the quantum gravitational corrections.  Using this equilibrium description, the entropy of black holes scales with the area of their event horizon, while their temperature scales with the surface gravity \cite{1c, 1a0}.   This approximation holds only for sufficiently large black holes. At these larger scales, the temperature is exceedingly low, allowing us to disregard thermal fluctuations and use equilibrium thermodynamics. However, as black holes reduce in size due to Hawking radiation  \cite{Kanti:2014dxa,Pappas:2016ovo,Sakalli:2022xrb}, quantum gravitational corrections can no longer be neglected.   Similarly, at such smaller scales, the temperature rises and thermal fluctuations cannot be dismissed. These fluctuations can be explored as perturbative corrections to equilibrium thermodynamics, leading to logarithmic corrections to the entropy of a black hole \cite{32, 32a, 32b, 32c, 32d}. Furthermore, it is known that the geometry can be derived from thermodynamics using the Jacobson formalism \cite{gr12}.  Therefore, quantum fluctuations in geometry can be connected to thermal fluctuations in equilibrium thermodynamics using the Jacobson formalism \cite{gr14}. This insight has spurred research into quantum gravitational corrections for various black hole scenarios \cite{12P, 12ap, 13p, 19p, 4p, hbh}. Additionally, the holographic principle \cite{4a, 5a} has been used to analyze quantum gravitational corrections in black hole thermodynamics \cite{6a, 7a, 18, 18a, 18b, 18c, 18d}. This was achieved by investigating the back reaction of the geometry using the finite $N$ limit of boundary conformal field theory. These corrections correspond to $\alpha'$ corrections in the geometry, which, in turn, give rise to higher curvature corrections. These higher curvature corrections are known to introduce modifications to the standard thermodynamics of black holes \cite{wald1, wald2}. Apart from these approaches, other approaches have also been used to obtain corrections to black hole thermodynamics \cite{6a, 7a, 18, 18a, 18b, 18c, 18d}. 
These corrections correspond only to perturbative quantum gravitational corrections to geometry, which are obtained from perturbative thermal corrections to equilibrium thermodynamics. These corrections occur at a scale so small that neglecting the perturbative corrections to equilibrium thermodynamics is not permissible. However, this scale is still much larger than the Planck scale, near which the full non-perturbative quantum gravitational corrections become significant, making it impossible to analyze the geometry using only perturbative quantum gravitational corrections.

Non-perturbative quantum gravitational corrections to the thermodynamics of various black holes have been investigated \cite{6ab, 7ab, 7ba}, leading to the observation that such corrections can introduce non-trivial modifications to the thermodynamics of black holes. We would like to clarify that such non-perturbative corrections become important at a scale, when purely quantum gravitational effects become significant, but are still analyzed at a scale above the Planck scale.  So, these non-perturbative corrections are assumed to occur at a scale above but close to the Planck scale.
Such non-perturbative quantum gravitational corrections to the thermodynamics of black holes have also been derived using string theoretical effects \cite{Dabholkar, ds12, ds14}. Additionally, general arguments based on the properties of conformal field theories have been used to obtain such corrections \cite{main}.   
The influence of full non-perturbative quantum gravitational corrections on the thermodynamic behavior of small four-dimensional Schwarzschild black holes \cite{bia}, Born–Infeld black holes \cite{bia1}, AdS black holes \cite{j1}, Myers-Perry black holes \cite{j2}, and a system of M2-M5 branes \cite{j6} have been investigated. These non-perturbative corrections introduce an exponential term to the black hole entropy, consequently modifying other thermodynamic quantities. Furthermore, these corrections may play a crucial role in the short-distance stability analysis of quantum-scale black holes, affecting their heat capacity, which, in turn, determines their stability. This has significant implications for black hole evaporation, including the black hole information paradox \cite{Mathur:2009hf,Almheiri:2019psf,Harlow:2014yka,Preskill:1992tc,Sakalli:2012zy,Sakalli:2010yy}. Thus, it becomes important to properly classify the effects of such corrections at various scales and in various dimensions, which we accomplish in this paper. However, spacetime is restricted to scales above the Planck scale, as the black hole thermodynamics and even the spacetime structure cannot be defined below the Planck scale.

We will explicitly use a quantum-corrected metric to perform this analysis.  Here, the geometry is replaced by a quantum geometry to account for quantum gravitational corrections, and this has been thoroughly studied for various approaches to quantum gravity. This quantum-corrected metric can be expressed in terms of the classical metric and quantum corrections to it.   The quantum Raychaudhuri equation has been used to construct such a quantum metric, and it has been demonstrated that such a quantum metric changes the Hawking radiation and black hole thermodynamics significantly near the Planck scale \cite{qg1, qg2}. Another proposal to construct quantum metric to account for quantum gravitational corrections has led to the development of rainbow gravity \cite{qg3}, and here again it has been demonstrated that the quantum corrected metric modifies the thermodynamics of black hole near Planck scale \cite{qg4}.  
The corrections to Hawking radiation near the Planck scale in noncommutative geometry have also been studied using this approach \cite{qg5}. Finally, T-duality in string theory has also been used to study the quantum corrections to a black hole metric, and its implication on the thermodynamics near the Planck scale \cite{qg6}. Using Jacobson formalism \cite{gr12}, it has been demonstrated that corrections to the entropy of a black hole can be directly used to construct a quantum-corrected metric for it \cite{gr14}. 
This construction of a quantum-metric using  
Jacobson formalism has also been generalized to black branes \cite{b1}.
Thus, motivated by these works, we will also use a quantum-corrected metric to analyze quantum gravitational corrections to the geometry. 

{It is well known that such quantum correction first introduces logarithmic corrections to the entropy of a black hole \cite{32, 32a, 32b, 32c, 32d, SPR}. Then at very small lengths scales (near Planck length $L_P$), the corrections to the black hole entropy can be expressed as an exponential function \cite{Dabholkar, ds12, ds14, main}. We will use this exponentially corrected entropy for our analysis.  As spacetime starts to break down below this scale \cite{min1, min2, min3, min4, mi}, geometry (including black hole geometry) cannot be defined below this scale $L_{min}$. This brings us to the main physical result of this paper. Even though the breaking of spacetime is a universal consequence of quantum gravity \cite{Das:2010sj}, the precise scale at which it breaks down is based on heuristic arguments \cite{h1, h2}. So, the identification  $L_{min} =L_P$ is only based on heuristic arguments.  In fact,  it has also been argued that spacetime metrics might break down above the Planck scale ($L_{min}> L_P$) \cite{Das:2010sj}. Such breaking of spacetime above the Planck scale can also have consequences for quantum mechanics, which could be detected through ultra-sensitive measurements \cite{hn, Pikovski:2011zk}. Such back-reaction of Planckian quantum gravitational corrections to quantum mechanics have been studied for various quantum systems, such as anomalous moment for the muon  \cite{Das:2011tq}, optomechanical systems \cite{Bosso:2016ycv},   Lamb shift,  Landau levels, and even a minute but detectable effect on the tunneling current in a scanning tunneling microscope \cite{Ali:2011fa},  transition rate of ultra-cold neutrons in gravitational field \cite{Pedram:2011xj}, and several other systems \cite{Bosso:2023aht}. Such modifications to black hole physics have also been studied, and specifically, the back-reaction of quantum gravitational corrections to the  Hawking radiation has been investigated \cite{1, 2, 3, 4,5,6}.   These quantum gravitational corrections only modify the Hawking radiation near the Planck scale, and we obtain the standard Hawking radiation for large black holes, where such corrections can be neglected. The quantum gravitational corrections to the Parikh-Wilczek have also been studied   \cite{Nozari:2009nr,Nozari:2012nf,Chen:2013ssa, Arzano:2005rs}. Furthermore, as the scale at which spacetime breaks down is based on heuristic argument \cite{h1, h2}, it is possible to propose that quantum gravitational effects break the spacetime metric at length scales below Planck scale ($L_{min}< L_P$), and thus forming sub-Planckian black holes \cite{Carr:2015nqa, Ling:2021olm}. If we properly analyze the breaking of spacetime, we would not find a sharp end to spacetime, but at a scale ($L_{min}) $ around the Planck scale, the structure of spacetime metric would slowly become fuzzy, and thus lose meaning. Thus, if we analyze black holes at smaller scales, we would initially gain information about them. However, as the spacetime starts to become fuzzy due to quantum fluctuations, we would observe that   information about this system would approach an constant. It would not be possible to obtain further information below this scale. This scale at which we cannot gain any further information about the system due to a breaking  of  spacetime would occur near the Planck scale, but can in principle occur about or below it. This behavior can be precisely quantified using Fisher information. Thus, we will use Fisher information \cite{fish1, fish2} to explicitly analyze to what scale even a quantum corrected metric cannot obtain new Fisher information  about quantum gravitational corrections, as the spacetime breaks down around  that scale. }

\section{Quantum Gravitational Corrected Geometry}\label{sec2}
 In this section, we will analyze the effects of quantum gravitational corrections on the geometry of a higher-dimensional Schwarzschild black hole. We begin by first reviewing the properties of a higher-dimensional black hole. The metric of $d$-dimensional Schwarzschild black hole can be expressed as follows \cite{SPR, SPR2}:
\begin{eqnarray}\label{metric}
ds^{2}=-f(r)dt^{2}+\frac{dr^{2}}{f(r)}+r^{2}d\Omega_{d-2}^{2},
\end{eqnarray}
where the metric function $f(r)$ is given by the equation,
\begin{eqnarray}
f(r)=1-\frac{16\pi G_dM}{(d-2)\Omega_{d-2}r^{d-3}}.
\end{eqnarray}
Here, $G_d$ represents the  $d$-dimensional Newton's constant, and $M$ represents the black hole mass. Additionally, $d\Omega_{d-2}^{2}$ represents the metric of the ${d-2}$-dimensional unit sphere with an area of $\Omega_{d-2}=2{\pi}^{\frac{d-1}{2}} / \Gamma\left({\frac{d-1}{2}}\right)$. Using this metric, the radius of the horizon can be determined by the condition $f(r=r_{0})=0$ and can be explicitly expressed as $r_{0}=[ 16\pi G_dM /( (d-2)\Omega_{d-2})]^{\frac{1}{(d-3)}}$. 
Substituting this expression for the radius of the horizon into $S_d$, the original equilibrium entropy of the black hole can be written as \cite{Argyres}
\begin{equation}\label{entropy1}
S_d=\int \frac{1}{T(M)} \, dM = \frac{4\pi }{(d-2)} \left(\frac{16 \pi}{(d-2) M_p^{d-2}}\right)^{\frac{1}{d-3}} \left(\frac{\Gamma\left(\frac{d-1}{2}\right)}{2\pi^{(d-1)/2}}\right)^{\frac{1}{d-3}} \cdot M^{\frac{d-2}{d-3}} 
\end{equation}
Hawking temperature $T_0$ of the $d$-dimensional Schwarzschild black hole is given by $T_{0}=(d-3)/4\pi r_{0}$. It has been argued that non-perturbative quantum gravitational effects produce an exponential correction to the equilibrium entropy of the black hole, such that  \cite{main, Dabholkar, ds12, ds14, exp12,  exp14} 
\begin{eqnarray}\label{corrected}
S_Q = S_d + \eta \ e^{-S_d},
\end{eqnarray}
where  $S_Q$ is the quantum gravitationally corrected entropy of the black hole and $\eta$ is a parameter. Here we would like to clarify that for large black holes, when even quantum gravitational corrections can be neglected, the entropy is given by $S_d $, and $\eta =0$. However, for smaller black holes, where leading order perturbative quantum gravitational corrections cannot be neglected, the corrected entropy is given by $S_{Q1} = \alpha_1\log A$, where $\alpha_1$ is a parameter and $A$ is the area of the horizon \cite{32, 32a, 32b, 32c, 32d}. The next-to-the-leading order corrections are given by $S_{Q2} = \alpha_2/A$, where again $\alpha_2$ is another parameter \cite{hi12, hi14}. These parameters depend on the details of the theory to obtain the quantum gravitational corrections.  However,  as the black hole becomes very small (close to but above the $L_{min}$), these perturbative corrections cannot be used, and the full non-perturbative corrections have to be used. Such non-perturbative corrections modify the original expression for the black hole entropy to  
$S_Q =S_d + \eta \ e^{-S_d}$  \cite{main, Dabholkar, ds12, ds14, exp12,  exp14}. Here, like the value of parameters used in perturbative quantum corrections, i.e. $\alpha_1, \alpha_2$, this parameter obtained 
 from the non-perturbative quantum gravitational corrections also depends on the details of the theory  \cite{bia, bia1, j1, j2}. For large black holes,  $\eta = 0$,  and for corrections obtained from string theory   $\eta = 1$  \cite{Dabholkar, ds12, ds14}. However, since the parameter governing quantum gravitational corrections to the black hole depends on the details of the approach \cite{12P, 12ap, 13p, 19p, 4p, hbh}, we shall treat $\eta$ as a general parameter and analyze the behavior of the black hole for various values of $\eta$ \cite{bia, bia1, j1, j2}.

We can now utilize this quantum gravitationally corrected entropy to analyze the effects of quantum gravitational corrections on the geometry of a $d$-dimensional Schwarzschild black hole. It is well-established that this geometry can be derived through thermodynamics in the Jacobson formalism \cite{gr12}. Consequently, at short distances, quantum fluctuations in the geometry can be linked to thermal fluctuations in the thermodynamics of the black hole  \cite{gr14}. Therefore, it becomes possible to use corrections to the equilibrium entropy of a black hole to derive quantum gravitational corrections to its metric. This approach is designed to ensure that the quantum-corrected metric directly yields the quantum gravitational corrected entropy. In pursuit of this objective, we introduce a modified metric for a Schwarzschild black hole, which naturally generates the corrected entropy in an exponential form. 
The idea of replacing geometry, by a quantum geometry to account for quantum gravitational corrections has been thoroughly studied for various approaches to quantum gravity \cite{qg1, qg2, qg3, qg4, qg5, qg6}. In fact,  motivated by Jacobson formalism \cite{gr12}, it has been argued  that corrections to the entropy of a black hole  can be directly used to construct a quantum-corrected metric for it \cite{gr14}. This is possible as according to Jacobson formalism \cite{gr12}, geometry of spacetime emerges from thermodynamics. Such construction of a quantum-metric using Jacobson formalism has also been done for black branes  \cite{b1}.  Thus, motivated the construction of quantum corrected metric used in various different approaches to quantum gravity \cite{qg1, qg2, qg3, qg4, qg5, qg6}, and the Jacobson formalism \cite{gr12}, we will use the exponential corrections to the entropy  \cite{main, Dabholkar, ds12, ds14, exp12,  exp14}  to construct a quantum-corrected metric.

Now we use the Jacobson formalism (see appendix) to explicitly analyze the quantum gravitationally corrections to geometry due to quantum gravitationally corrected entropy.  In fact, it has already been demonstrated that a quantum metric is obtained from the Jacobson formalism by using   quantum gravitationally corrected entropy  \cite{gr14}. Here, we will use this formalism for a 
 $d$-dimensional Schwarzschild. 
So, this  framework will enables us to derive quantum corrected  metric,  that incorporate quantum corrections to $d$-dimensional Schwarzschild black hole entropy. This metric can be written as 
\begin{eqnarray}\label{metric1}
ds^{2}=-f(r)_{Q}dt^{2}+\frac{dr^{2}}{f(r)_{Q}}+h(r)_{Q}d\Omega_{d-2}^{2},
\end{eqnarray}
where $f(r)_{Q}$ and $h(r)_{Q}$ are corrected metric functions.
Now we choose,
\begin{equation}\label{hQ}
h(r)_{Q}=\frac{1}{\Omega_{d-2}}\left[4G_d(S_d + \eta \ e^{-S_d})\right]^{\frac{2}{d-2}},
\end{equation}
and define  a quantum gravitationally corrected area $A_{Q}$, such that  it reproduce the quantum gravitationally corrected entropy as $S_{Q}={A_{Q}}/{4G_{d}}$ (\ref{corrected}). The metric (\ref{metric1}) is constructed so that the standard entropy obtained from this quantum gravitationally corrected metric is the quantum gravitationally corrected entropy of a $d$-dimensional Schwarzschild black hole given in Eq. (\ref{corrected}). It is easy to see that if we set $\eta=0$, then the original entropy (\ref{entropy1}) is recovered.  Now this quantum gravitationally corrected effective metric can also be used to investigate the effects of quantum gravitational corrections on various thermodynamic quantities. Consequently, we observe that the temperature is modified due to quantum gravitational effects as,
\begin{equation}\label{fQ0}
T_{Q}=\frac{dM}{dS_{Q}}=\frac{f^{\prime}(r_{0,Q})_Q}{4\pi},
\end{equation}
where $r_{0,Q}$ is the root of $f(r= r_{0, Q})_Q=0$. We will now obtain an explicit expression for  $f(r)_{Q}$. In order to do that we assume it in the following form,
\begin{equation}\label{fQ}
f(r)_{Q}=f(r)+\eta g(r),
\end{equation}
where $f(r)$ is given by the Eq.  (\ref{metric1}) and $g(r)$ is an unknown function which should be determined.
Now using  $S_{Q}$ in the first law of thermodynamics ($T_{Q}={dM}/{dS_{Q}}$), we observe
\begin{equation}\label{TQ}
T_{Q} = T_{0} + \eta T_{0} e^{-S_{d}},
\end{equation}
where $T_{0}={dM}/{dS_{d}}$.
Putting the Eq. (\ref{fQ}) into the Eq. (\ref{fQ0}) and compare the result with the Eq. (\ref{TQ}), we obtain  
$
\left(g^{\prime}_Q\right)_{r= r_{0, Q}}=4\pi T_{0} e^{-S_{d}}=(d-3)\frac{e^{-S_{d}}}{r_{0}}.
 $
Now, we can express $g(r)$ as following,
\begin{equation}\label{g}
g(r)=-\frac{4 G_{d} (d-3)}{(d-2)\Omega_{d-2} r_{0}^{d-2}}\exp\left({-\frac{\Omega_{d-2}r^{d-2}}{4G_d}}\right).
\end{equation}
Using this solution in the Eq. (\ref{fQ}), we  obtain,
\begin{equation}\label{fQr}
f(r)_{Q}=1-\frac{16\pi G_dM}{(d-2)\Omega_{d-2}r^{d-3}}-\eta \frac{4 G_{d} (d-3)}{(d-2)\Omega_{d-2} r_{0}^{d-2}}\exp\left({-\frac{\Omega_{d-2}r^{d-2}}{4G_d}}\right).
\end{equation}
Using the quantum metric functions  given by Eqs. (\ref{fQr}) and (\ref{hQ}), we constructed the quantum correction to the metric function given in Eq. (\ref{metric1}).
Now quantum corrected  horizon radius $r_{0, Q}$ is obtained by $f(r)_{Q}=0$. This can be used to obtain  a  quantum correct mass for the black hole as,
\begin{equation}\label{correctedM}
M_Q = \frac{(d-2)\Omega_{d-2}r_{0}^{d-3}}{16\pi G_d}\left[1-\frac{4 G_{d} (d-3) \eta}{(d-2)\Omega_{d-2} r_{0}^{d-2}}\exp\left({-\frac{\Omega_{d-2}r_{0}^{d-2}}{4G_d}}\right)\right] = M\left[1-\eta C M^{\frac{2-d}{d-3}} e^{-S_d}\right],
\end{equation}
where $C=[({4 G_{d} (d-3)})/({(d-2)\Omega_{d-2})}][({(d-2)\Omega_{d-2}})/({16\pi G_{d}})]^{\frac{d-2}{d-3}}$.
Here, the  original mass of this black hole   $M={(d-2)\Omega_{d-2}r_{0}^{d-3}}/{16\pi G_d}$ is obtained by setting the Eq. (\ref{metric1}) equal to zero at $r=r_{0}$. 
It can be observed that the first law of black hole thermodynamics is satisfied by these quantum gravitationally corrected expressions. 
Now the main aim of the quantum corrected mass given in Eq. (\ref{correctedM}) is to act as a physical quantity that can  quantify the effects of quantum gravitational effects on important information theoretical aspects of this  black hole.  This will be done in the next section.  
\section{Kullback-Leibler divergence} \label{sec4}
In the previous sections, we analyzed the modifications to a higher-dimensional black hole from quantum gravitational corrections. Now, we quantify the deviations from the standard behavior due to such corrections. This will be done using Kullback-Leibler divergence \cite{kl1, kl2}, which measures how much two probability distributions differ from each other. Even though Kullback-Leibler divergence is not symmetric, and hence not a metric, it does give a realistic estimation of how far a given probability distribution is from another probability distribution. It is possible to obtain the probability distribution of the particles emitted from a black hole during its evaporation using the Parikh-Wilczek formalism \cite{pw, pw1}, which considers the back reaction of the emitted particles on the black hole geometry and proposes that the entropy of the black hole can be expressed in terms of the probability distribution of the emitted particles. 
As the entropy of the black hole is modified by quantum gravitational corrections, this probability distribution in the Parikh-Wilczek formalism will also be modified. We will explicitly analyze such modifications to the probability distribution of the emitted particles, and use them to estimate the deviation of the behavior of a quantum-corrected black hole from the original black hole. 
For a higher dimensional Schwarzschild black hole, we assume that  $P_n$  is the probability that the black hole evaporates by radiating  $n$ particle.  
We can write this probability as 
\begin{equation}
 P_n(0)=\frac{\Omega_n}{\Omega_{total}},
\end{equation}
where $\Omega_n$ is the number of microstates when a particle evaporates into $n$ particle, and $\Omega_{total}=\sum_1^{\infty}\Omega_n $ is the total number of the possible microstates of the system. It has been demonstrated that for a Schwarzschild black hole of mass $M$, $P_n$ can be expressed in terms of $M$ \cite{pw1}. Although the expression was explicitly analyzed for a four-dimensional black hole, the arguments used to obtain this distribution \cite{pw1} are very general and hold in any dimension. 
The higher dimension gravitational constant $G_d$ is related to the Planck mass $M_P$ as 
$G_d = M_P^{2-d}$. To simplify the expressions,  we  define a new constant as 
\begin{equation}
     \mu_d =  \frac{1}{\sqrt{\pi}(d-2)} \left( \left( \frac{8\Gamma\left(\frac{d-1}{2}\right)}{d-2} \right)^{\frac{1}{d-2}} \frac{1}{M_P} \right)^{\frac{d-2}{d-3}}
\end{equation}

The probability distribution $P_n$ is normalized. As the black hole evaporates by emitting  $n$  particles, the probability that one among $\Omega_n$ microstates occurs is $q_{\alpha}$. Since it is assumed that all of the $\Omega_{total}$ microstates occur with the same probability, we can write $q_\alpha=1/\Omega_n$. The total probability distribution for both the probabilities of the number of particles that a four-dimensional Schwarzschild black hole emits during its evaporation and the possible microstates has been obtained \cite{pw1}. This can be easily generalized to higher dimensions, where this  probability distribution can be written as 
\begin{equation}
    P_n(0) = q_\alpha \times \Omega_{total}=\frac{(4\pi \mu_d M^{\frac{d-2}{d-3}})^{n-1}}{(n-1)!}e^{-4\pi \mu_d M^{\frac{d-2}{d-3}}}= \frac{S_d^{n-1}}{(n-1)!} \cdot e^{-S_d},
\end{equation}
where the zero in $P_n(0)$ denotes the absence of any quantum gravitational corrections.  
Here, we observe that this probability distribution depends on the mass of the black hole. 

As the arguments are general, these also hold for a quantum-sized black hole, with the mass being replaced by the novel quantum mass of the black hole. This is because the quantum mass naturally reproduces the quantum-corrected entropy of the quantum-sized black hole. As the entropy obtained from the Parikh-Wilczek formalism \cite{pw, pw1} has to be consistent with the entropy of the black hole obtained from standard methods, the quantum gravitationally corrections to the entropy produced by the Parikh-Wilczek formalism \cite{pw, pw1} should also coincide with such corrections to the entropy obtained using standard methods. Thus, we can also write the probability distribution for a quantum-corrected higher-dimensional Schwarzschild black hole, by replacing the usual mass $M $ with the novel quantum mass  $M_Q = M\left(1-\eta C M^{\frac{2-d}{d-3}} e^{-S_{d}}\right)$. This replacement is done as the novel quantum mass was constructed so that the entropy obtained from it using standard methods is the quantum-corrected entropy. Thus, using the novel quantum mass, we obtain the 
quantum-corrected probability distribution as 
\begin{equation}
    {P}_{n} (\eta) = \frac{(4\pi \mu_d M_Q^{\frac{d-2}{d-3}})^{n-1}}{(n-1)!} e^{-4\pi \mu_d M_Q^{\frac{d-2}{d-3}}} 
    = \frac{{S_Q}^{n-1}}{(n-1)!} e^{-S_Q}
\end{equation}

Here $S_Q$ is the entropy of the corrected black hole. 
Now using $ {P}_{n} (\eta)|_{\eta =0} = P_n (0)$, we can calculate the  Kullback-Leibler divergence between the original and quantum gravitationally corrected probability distributions. 
Thus, we now use the  Kullback-Leibler divergence \cite{kl1, kl2} to measure deviations from quantum gravitational corrections.  The Kullback-Leibler divergence is given by 
\begin{eqnarray}
  D_{KL}(  {P}_{n} (\eta) || P_n (0))  
  &=& \sum_{n=1}^{\infty} P'_n \left( (n-1) \log\left( \frac{S_Q}{S_d} \right) - (S_Q - S_d) \right) \nonumber \\
  &=& \log\left( \frac{S_Q}{S_d} \right) \sum_{n=1}^{\infty} P'_n (n-1) - (S_Q - S_d) \sum_{n=1}^{\infty} P'_n \nonumber \\
  &=& S_Q \log \left( \frac{S_Q}{S_d} \right) - (S_Q - S_d) \nonumber \\
  &=& 4\pi \mu_d M_Q^{\frac{d-2}{d-3}} \frac{d-2}{d-3} \log \left( \frac{M_Q}{M} \right) - 4\pi \mu_d \left( M_Q^{\frac{d-2}{d-3}} - M^{\frac{d-2}{d-3}} \right)
\end{eqnarray}
\begin{figure}
    \centering
    \begin{subfigure}[H]{0.7\textwidth}
        \centering
    \includegraphics[width=\linewidth]{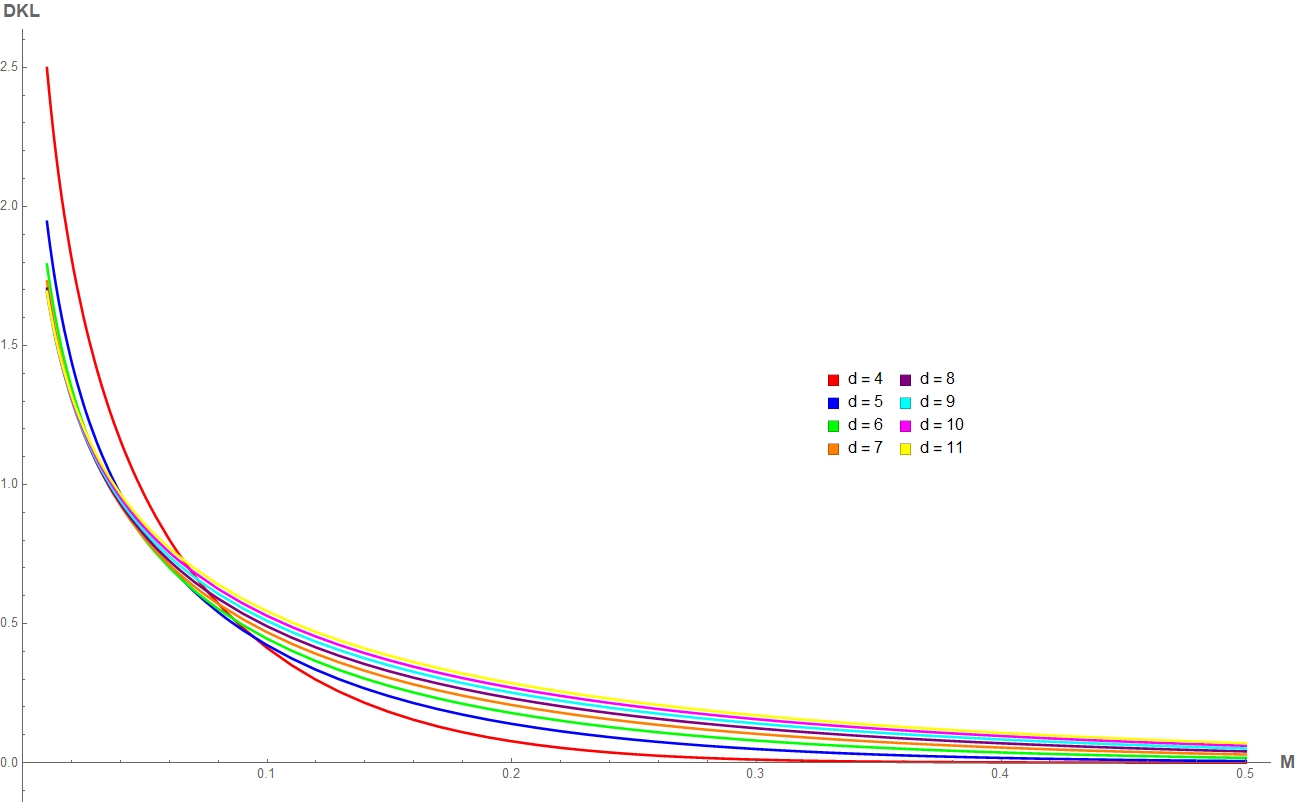}
        \caption{Plot for $D_{KL}$  at a fixed value of $\eta= 0.5$.}
        \label{fig:DKL2D}
    \end{subfigure}
    \hfill
    \begin{subfigure}[b]{0.9\textwidth}
        \centering
        \includegraphics[width=\linewidth]{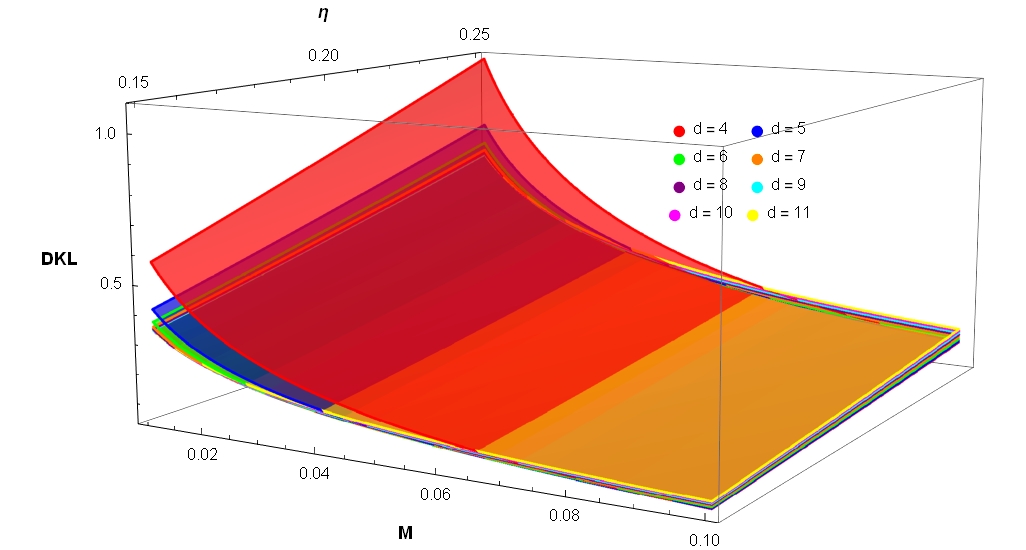}
        \caption{Plot for $D_{KL}$ for various values of $\eta$.}
        \label{fig:DKL3D}
    \end{subfigure}
    \caption{ Plots  for $D_{KL}$ in various different dimensions.}
    \label{fig:Figure1}
\end{figure}


\begin{figure}
    \centering
    \begin{subfigure}[b]{0.7\textwidth}
        \centering
        \includegraphics[width=\linewidth]{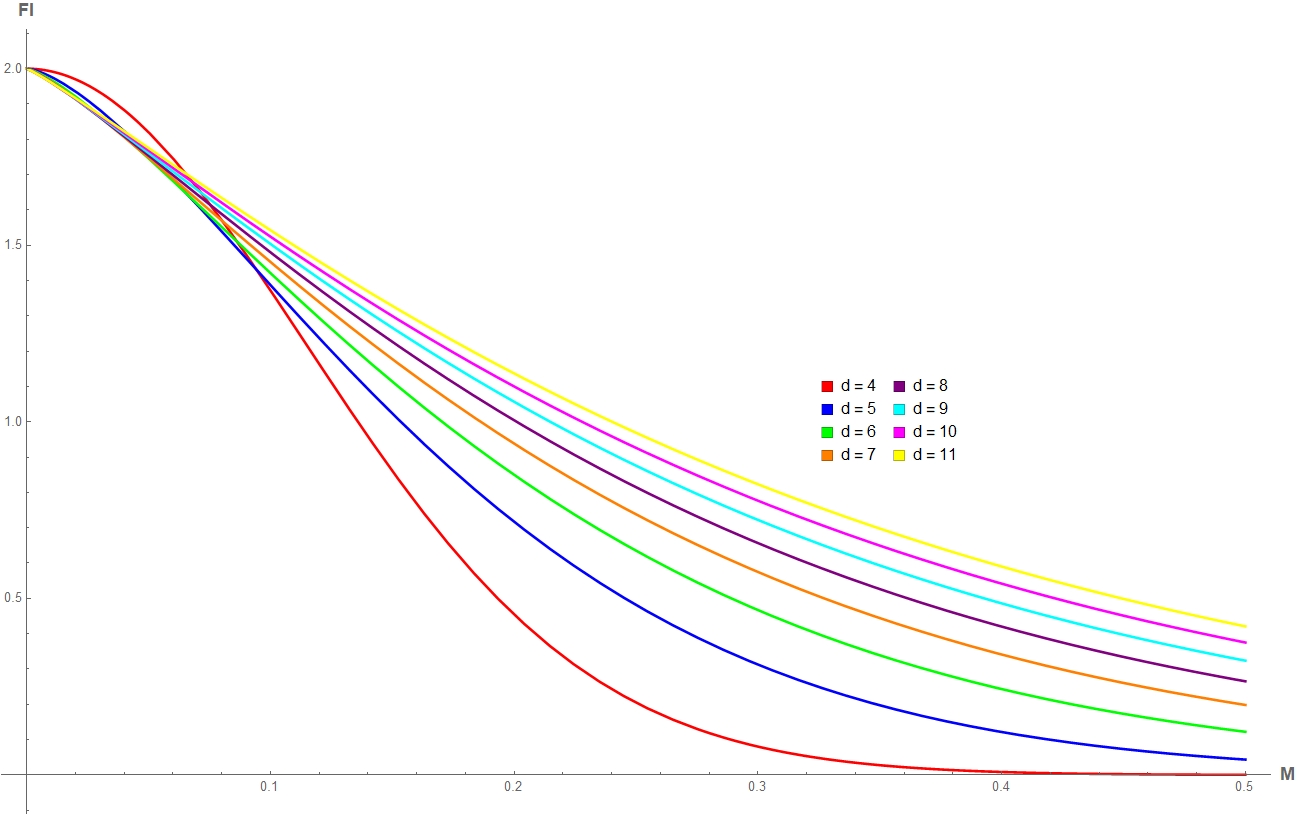}
        \caption{  Plot for $FI$ for a fixed value of $\eta= 0.5$.}
        \label{fig:FI2D}
    \end{subfigure}
    \hfill
    \begin{subfigure}[b]{0.8\textwidth}
        \centering
        \includegraphics[width=\linewidth]{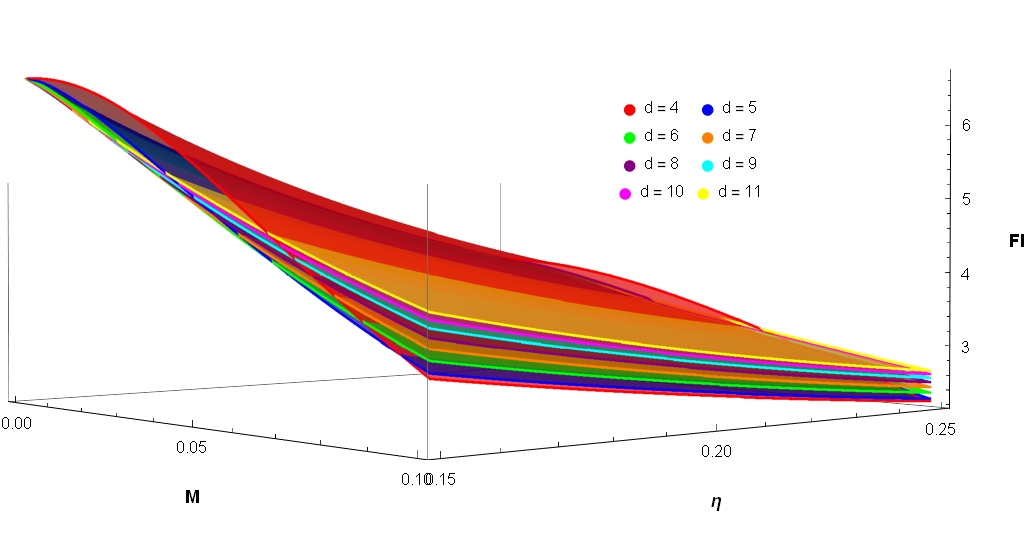}
        \caption{Plot for $FI$ for different values of $\eta$.}
        \label{fig:FI3D}
    \end{subfigure}
    \caption{ Plots $FI$ for different dimensions.}
    \label{fig:Figure2}
\end{figure}


We approach this problem by treating it as a black-box analysis, focusing on the particles emitted from an evaporating black hole as it decreases in size. The probability distribution of these emitted particles is influenced by quantum gravitational effects, which can be measured at a distance from the black hole itself. This methodology allows us to capture the behavior of the black hole near the Planck scale.
To quantify the quantum gravitational corrections, we utilize the Kullback-Leibler  divergence.
We first observe from Fig. (\ref{fig:Figure1}) that the $D_{KL}$ increases with decreases in scale. We also observe that it depends on the dimensions. For large black holes, the $D_{KL}$ is maximum for higher dimensional black holes, however, after Planck scale $D_{KL}$ is maximum for lower dimensional black holes. This indicates that some kind of new physics occurs near Planck scale. This is expected as quantum gravitational effects are expected to break the spacetime down around this scale. 
However, to properly analyze the effects of quantum gravitational corrections, we will have to use Fisher information.

\section{Fisher Information}
In the previous section, we analyzed the Kullback-Leibler divergence between the original and corrected probability distributions and observed that its behavior changed near critical value near Planck scale. In this section, we will properly analyze this by using the concept of Fisher information. Fisher information measures the information about a parameter that can be obtained from a probability distribution \cite{fish1, fish2}. To analyze how much information can be obtained about quantum gravitational corrections from the probability distribution of particles emitted from a black hole during its evaporation, we obtain Fisher information of the parameter $\eta$. In other words, we are investigating how the information about quantum gravitational corrections depends on the scale. 
As Fisher information is related to Kullback-Leibler divergence \cite{fish4, fish5}, we measure Fisher information by first defining Kullback-Leibler divergence between two corrected probability distributions over different values of $\eta$; namely: $P_{n} (\eta)$ and $P_{n} (\eta +\delta \eta)$.  We measure the Fisher information associated with $\eta$, as we move in the $\eta$ space between $\eta$ and $\eta +\delta \eta$. In the limiting case, we will get our previous results back, when we set $\eta =0$. 
Now we can use the Taylor expansion to expand $  P_{n} (\eta + \delta \eta) $, and find its relation to $  P_{n} (\eta) $
\begin{equation}
  P_{n} (\eta + \delta \eta)  \approx P_{n} (\eta )+ \delta \eta \frac{\partial P_{n} (\eta )}{\partial \eta}=P_{n} (\eta )+\delta P_{n}  (\eta ).
\end{equation}
Here, we define the correction to the probability distribution from $\eta$ to $\eta + \delta \eta $ as $\delta P_{n}  (\eta )$. 
 We can write the Kullback-Leibler divergence between the probability distributions at $\eta$ and $\eta + \delta \eta$ as  
\begin{eqnarray}
     D_{KL}(P_{n} (\eta + \delta \eta)||P_{n} (\eta)) &=& \sum^{\Omega}_{n =1}
    (P_{n} (\eta +\delta \eta )\log \frac{P_{n} (\eta +\delta \eta )}{P_{n} (\eta )}).
    \end{eqnarray}
   We write the argument of the logarithm  as $(1+\delta P_{n}(\eta)/P_{n}(\eta))$, and thus obtain 
   \begin{eqnarray}
 D_{KL}(P_{n} (\eta + \delta \eta)||P_{n} (\eta)) &=&  \sum^{\Omega}_{n =1} \left(P_{n} (\eta )+\delta P_{n} (\eta) \log\frac{P_{n} (\eta )+\delta P_{n} (\eta)}{P_{n} (\eta )}\right)  \nonumber \\ 
 &=& \sum^{\Omega}_{n =1} \left( P_{n} (\eta) \log \left(1+\frac{\delta P_{n} (\eta)}{P_{n}(\eta)}\right)\right ) +\sum^{\Omega}_{n =1}     
  \log\left(1+\frac{\delta P_{n}(\eta)}{ P_{n}(\eta)})
  \delta P_{n}(\eta)\right).   
    \end{eqnarray}
   We expand the logarithm term up to the second order in $\delta P_n $, and neglect other higher order terms. Thus, 
 using $\log(1+{\delta P_{n} (\eta)}/{P_n (\eta)})\approx ({\delta   P_{n} (\eta)}/{ P_{n} (\eta)})-({(\delta   P_{n} (\eta))^2}/{2  P_{n} (\eta)^2}) $, we  write 
   \begin{eqnarray}  D_{KL}(P_{n} (\eta + \delta \eta)||P_{n} (\eta)) \approx   \sum^{\Omega}_{n =1} \delta 
     P_n (\eta) +\frac{1}{2} \sum^{\Omega}_{n =1}  \left( P_n (\eta)\frac{(\delta   P_{n} (\eta))^2}{ P_n (\eta)^2}\right). 
    \end{eqnarray}
We can use the expression for the probability distributions, and obtain an explicit expression for the Kullback-Leibler divergence in terms of the mass of the black hole.  Moreover, we express $\delta P_{n} (\eta )= \delta \eta {\partial P_{n} (\eta )}/{\partial \eta}$ , and $(\delta P_{n} (\eta ))^2=(\delta \eta)^2 ({\partial P_{n} (\eta )}/{\partial \eta})^2$. Putting these back in the expression above  we can express the Kullback-Leibler divergence as 
  \begin{eqnarray}
   D_{KL}(P_{n} (\eta + \delta \eta)||P_{n} (\eta) ) &\approx&  \sum^{\Omega}_{n =1} \delta \eta \frac{\partial P_{n} (\eta )}{\partial \eta} + \sum \frac{1}{2} (\delta \eta)^2 P_{n} (\eta )  \Bigg(\frac{1}{ P_{n} (\eta )}\frac{\partial P_{n} (\eta )}{\partial \eta}\Bigg)^2\nonumber
    \end{eqnarray}
    
Since Kullback-Leibler divergence has a minima at $\eta=0$,  the first derivative of Kullback-Leibler divergence  is  zero at $\eta=0$, 
\begin{equation}
   D_{KL}(P_{n} (\eta + \delta \eta)||P_{n} (\eta ))= \frac{1}{2} (\delta \eta )^2 {E}\left(\Bigg( \frac{\partial \log P(\eta )}{\partial \eta } \Bigg)^2\right)\Bigg|_{\eta=0}.
\end{equation}
Here $ {E} $ denotes the expectation value.
Thus, the  Fisher Information $FI$ can be obtained from the Kullback-Leibler divergence as 
\begin{equation}
    FI(\eta)\approx \frac{\partial^2}{\partial ^2\eta } D_{KL}(P_{n} (\eta + \delta \eta)||P_{n} (\eta) ) |_{\delta\eta =0}= {E}\Bigg( \frac{\partial \log P(\eta )}{\partial \eta } \Bigg)^2. 
\end{equation}  Now using the explicit expression for the Kullback-Leibler divergence between a corrected value and an original value at $\eta =0$, we can also 
 write an explicit expression for  the Fisher information 

\begin{eqnarray}
FI(\eta)=\mathbb{E} \left[ \left( \frac{\partial}{\partial \eta} \log P'(n) \right)^2 \right] 
\approx\mathbb{E} \left[ \left( \frac{\partial S_Q}{\partial \eta} \left( \frac{n-1}{S_Q} - 1 \right) \right)^2 \right] 
 \end{eqnarray}
Factor out \( \frac{\partial S_Q}{\partial \eta} \), we can write the Fisher information  
\begin{equation}
FI(\eta)\approx\left( \frac{\partial S_Q}{\partial \eta} \right)^2 \mathbb{E} \left[ \left( \frac{n-1}{S_Q} - 1 \right)^2 \right]
\end{equation}
Now we can write an expression for $S_Q$ as 
\begin{equation}
\mathbb{E} \left[ \left( \frac{n-1}{S_Q} - 1 \right)^2 \right] = \frac{1}{S_Q}
\end{equation}
Thus, the Fisher information with respect to \( \eta \) can be expressed as 
\begin{equation}
    FI(\eta)\approx\left( \frac{\partial S_Q}{\partial \eta} \right)^2 \cdot \frac{1}{S_Q}
\end{equation}

We observe from Fig. \eqref{fig:Figure2}, that the Fisher information about quantum gravitational corrections  decreases as the size of black holes become large, and can be neglected for  larger black holes. This is expected, as the quantum gravitational corrections only have negligible effects at larger scales.     We also observe from Fig. (\ref{fig:FI3D}) that for small black holes, where the  Fisher information cannot be neglected, it   is less for lower dimensional black holes than higher dimensional black holes.  However, around Planck scale, this behavior of  Fisher information changes, and it becomes more for lower dimensional black holes than higher dimensional black holes. Thus, again we observe a change in the behavior of Fisher information near Planck scale.  

Now we make an interesting observation. Using heuristic arguments it has been suggested that spacetime will break at the Planck scale $L_P$, and hence $L_P$  will act as a minimal length $L_{min}$ for spacetime.  However, this minimal  scale at which spacetime breaks has been assumed to be both above Planck scale $L_{min}> L_p$ \cite{1, 2, 3, 4,5,6}, and below it $L_{min}< L_p$ \cite{Carr:2015nqa, Ling:2021olm}. The important observation here is that no matter if $L_{min}$ is above, below or at Planck scale, there will be a bound to the Fisher information about quantum gravitational corrections. This is because it would not be possible to obtain information about quantum gravity beyond such a breaking of spacetime. Furthermore, we would also expect the breaking of spacetime to depend on the strength of quantum gravitational corrections. If these corrections are strong ($\eta$ is large), then spacetime would break faster as compered to  weak corrections ($\eta$ is small). This would imply that Fisher information should be inversely proportional to $\eta$. 

However, there is a problem with this heuristic arguments. We expect that quantum gravitational effects would slowly  break spacetime around Planck scale, rather than at a sharp point around Planck scale. The  spacetime becomes fuzzy around Planck  scale, and slowly become less well defined. Now the problem is to make this heuristic arguments rigorous. This can be done by addressing the problem from purely information theoretical perspective. Instead of analyzing the breaking of geometry, and the bounds it has on Fisher information, we can directly analyze the bounds on  Fisher information. If the Fisher information is bounded, then it is indicative of a  breaking of spacetime around Planck scale. 
We also observe that this is actually the case, and the  Fisher information about quantum gravity reaches a fixed value as the black hole evaporates, and this scales as $1/\eta$. It is interesting to note that this fixed value is inversely proportional to the strength of quantum gravitational corrections, as was physically expected from the heuristic arguments. Hence, we have been able to rigorously analyze the breaking of spacetime around the Planck scale, which had been  precisely heuristic argued. This was done by directly relating such breaking to the bounds on Fisher information about quantum gravitational effects.  So, this breaking limits the Fisher information about quantum gravity that can be obtained from the system. Hence, quantum gravity could restrict access to the information about itself, as such information here would be bounded by $1/\eta$.    

{The observation that the Fisher information about quantum gravitational effects reaches a fixed value during black hole evaporation, scaling as \(1/\eta\), where \(\eta\) quantifies the strength of quantum gravitational corrections, has profound implications for the black hole information paradox and even the  broader issue of meta-information. The fixed Fisher information value indicates a fundamental limit on the precision with which quantum gravitational corrections can be inferred from physical observables, such as the radiation emitted by an evaporating black hole. This inverse scaling with \(\eta\) is consistent with heuristic expectations, as stronger quantum gravitational effects impose stricter bounds on the retrievable information. Consequently, while quantum gravitational corrections may partially resolve the black hole information paradox by modifying the Hawking radiation spectrum to restore unitarity, they simultaneously restrict the accessibility of information about their own underlying framework.

This limitation introduces what can be described as a meta-information paradox, wherein not only is information about the black hole's initial state is lost, but information about the nature of quantum gravitational corrections themselves is intrinsically restricted. The bounded Fisher information suggests that quantum gravity corrections obscure aspects of their structure even as they modify the dynamics of black hole evaporation. 
The black hole information paradox has been addressed through various proposed solutions, most of which suggest that the paradox arises as an inferred  problem  at the black hole's horizon. These approaches argue that the paradox  will be resolved by a complete quantum theory of gravity in the ultraviolet  limit
\cite{Mathur:2009hf,Almheiri:2019psf,Harlow:2014yka,Preskill:1992tc,Sakalli:2012zy,Sakalli:2010yy, information, information12}.
In these proposals information paradox is expected to be solved in the  ultraviolet  limit. Thus, in these proposals it is suggested that  either information irretrievably lost and cannot be computationally recovered, or it is  not lost and can be  computationally recoverable.  Now this only possible if we can in principle have a full information about quantum gravitational corrections in the  ultraviolet  limit. 
However, we have demonstrated that quantum gravity prevents the full information about its own corrections in the ultraviolet limit. Thus, 
these quantum gravitational corrections obscure details about the fundamental quantum gravitational degrees of freedom. This reflects the theory's self-referential nature limits the  construction of a complete and consistent quantum theory of gravity. Such behavior bears resemblance to incompleteness theorems in formal logic and could, in principle, be connected to them \cite{in12, in14,int16}. Thus, according to this work, we cannot even address the information  paradox in the  ultraviolet  limit, as the meta-information about the formalism that can be used to address it is also obscured by its own self-referential nature. 

These results further suggest that the breaking of spacetime at the Planck scale, which is directly related to the bounds on Fisher information, limit the parameter space over which quantum gravitational effects can be discerned. This effectively restricts the distinguishability of competing quantum gravitational models from the observational data. For example, models such as string theory or loop quantum gravity, which predict deviations from classical physics at small scales, may become indistinguishable near Planck scale, due to the intrinsic boundedness of meta-information. The information associated with quantum gravitational corrections thus reaches a finite value, implying that even in principle, there is a limit to how much of the theoretical framework governing these effects can be obtained  from computational aspects of a physical systems.

This inherent boundedness has profound consequences for the study of black hole physics and quantum gravity. It imposes constraints on observational and computational approaches to resolving the black hole information paradox, suggesting that while it was suggested that the  information paradox could have been addressed by quantum gravitational corrections, our understanding of the corrections themselves remains fundamentally computationally incomplete. The implications extend beyond black hole thermodynamics, hinting at deeper properties of quantum gravity, such as its potential self-referential nature. In this view, quantum gravity  conceals aspects of its structure, reflecting a fundamental feature of the theory. }

\section{Stability at Quantum Scales} 
Understanding the stability of black holes at quantum scales is important for understanding  the interplay between quantum gravity and black hole thermodynamics. Classical black holes, such as Schwarzschild black holes, are thermodynamically unstable due to their negative specific heat, but quantum gravitational corrections can change this behavior. These corrections introduce parameters, such as \(\eta\), that capture the effects of quantum fluctuations on the black hole's thermodynamic quantities. By analyzing how \(\eta\) modifies the Helmholtz free energy, internal energy, and specific heat, we can explore new stability regimes and phase transitions that arise due to quantum effects. Moreover, the role of \(\eta\) can be understood through Fisher information, which provides a quantitative measure of the sensitivity of black hole stability to these quantum corrections. Thus, thermodynamic stability and information geometry can be related to each other, and this  offers a powerful tool for probing the quantum gravity.

\begin{figure}[h]
\resizebox{0.45\textwidth}{!}{%
\includegraphics{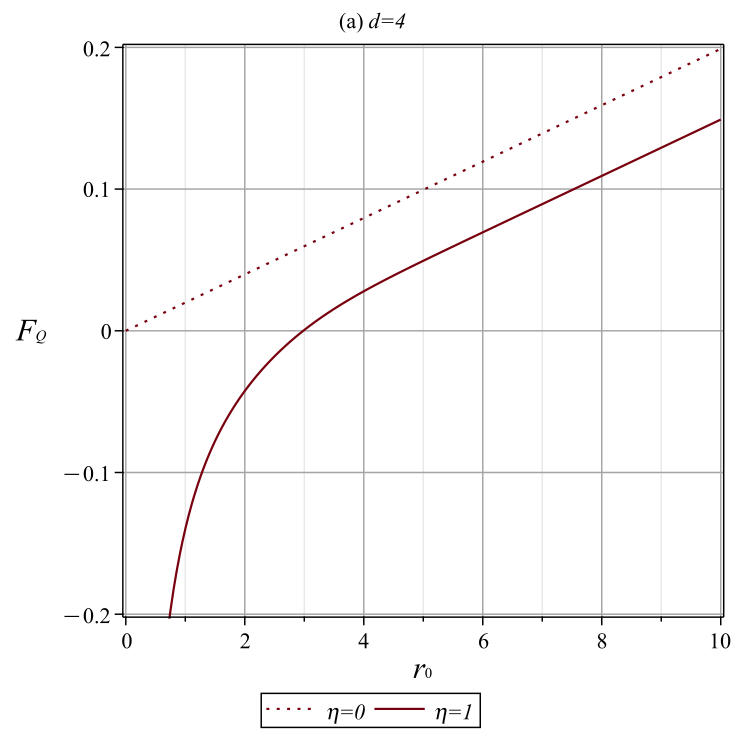}}\resizebox{0.45\textwidth}{!}{
  \includegraphics{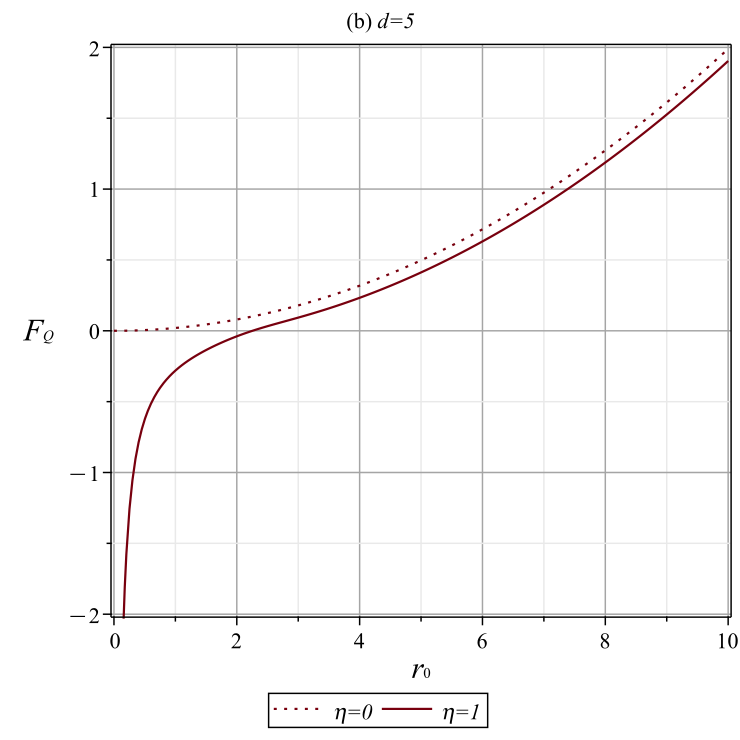}}
\vspace{0.5cm}       
\caption{Typical behavior of the $d$-dimensional Schwarzschild black hole Helmholtz free energy ($G_{d}=1$ is used).}
\label{figF}       
\end{figure}

So, as that quantum gravitational corrections may  change the stability at quantum scales, we will explicitly analyze them. This can be done by first using the quantum corrected metric to explicitly calculate the corrections to other thermodynamic quantities. 
Therefore, we can express the Helmholtz free energy of the black hole corrected for quantum gravitational effects as 
\begin{equation}\label{Helm1}
F_Q =-\int S_Q dT_{Q}=F_{0}+\eta F_{1},
\end{equation}
where $
F_{0}={\Omega_{d-2}}r_{0}^{d-3}/16 \pi G_d
 $
is uncorrected Helmholtz free energy. The corrected part $F_{1}$ is a complicated function of WhittakerM. In order to see the effects of correction we draw typical behavior of $F_{Q}$ for $d=4$ and $d=5$ in Fig. \ref{figF} (a) and (b) respectively. We can see that at large $r_{0}$ (large black hole) where quantum effects are negligible, then $F_{Q}\approx F_{0}$, however there are important deviation as small radii as expected. Opposite sign of the 
Helmholtz free energy at smaller $r_{0}$ may be sign of some important differences in large/small black hole stability. 

We can use the quantum gravitationally corrected Helmholtz free energy to obtain the expression for quantum gravitationally corrected internal energy
\begin{equation}\label{internal-S}
E_Q=F_Q+T_Q S_Q =E_{0}+\eta E_{1}=\frac{\Omega_{d-2}}{16 \pi G_d}(d-2)r_{0}^{d-3}+\eta E_{1},
\end{equation}
where $E_{1}$ is a complicated function of WhittakerM.

We observe that various thermodynamic quantities for a higher-dimensional black hole can also be directly obtained from the quantum gravitationally corrected metric. However, to analyze the stability of a quantum gravitationally corrected black hole, we use the quantum gravitationally corrected specific heat in terms of the original entropy as follows,
\begin{equation}\label{1111}
C_Q =T_{Q}\frac{dS_{Q}}{dT_Q}
= \frac{(d-2)S_{d}\left(\eta e^{-S_{d}}+1\right)\left(\eta e^{-S_{d}}-1\right)}{1+\eta(1+S_{d})e^{-S_{d}}}.
\end{equation}  
{We can see that at $\eta=0$ the uncorrected specific heat of the $d$-dimensional Schwarzschild black hole ($C_{0}=-(d-2)S_{d}$) is completely negative as well as 4-dimensional Schwarzschild black holes.}

 {In equation \ref{1111}, the numerator is positive, which means that to have a positive specific heat, the denominator should be positive. Using the condition for  thermodynamics stability, $C_Q\geq0$,  we obtain $\eta\geq e^{s_{d}}$. For a fixed mass of a black hole,  we can find the lower limit for $\eta$ for which the black hole is stable.  We know that the Schwarzschild black hole is unstable (at $\eta=0$), but in the presence of the exponential correction with the correction coefficient $\eta\geq e^{s_{d}}$, it may become stable at small radii. 
It is interesting to note that the  thermodynamics stability, $C_Q\geq0$,  can be related to $\eta$ as $\eta\geq e^{s_{d}}$. We note that  the Fisher information $FI(\eta)$ about $\eta$   is bounded by $1/\eta$. 
As quantum gravitational effects quantified by $\eta$ increases, the black hole becomes more thermodynamically stable. However, a more stable system responds less to external perturbations or fluctuations, reducing the information extractable from such changes. Fisher Information measures the sensitivity to measure quantum gravitational corrections. Now  the system is highly sensitive ($FI(\eta)$ is large), when the quantum gravitational effects are small, and so $\eta$ is small (and $1/\eta$ is large). Thus, there is a trade-off between Fisher information and stability. } 

\begin{figure}[tbh!]
\resizebox{0.45\textwidth}{!}{%
  \includegraphics{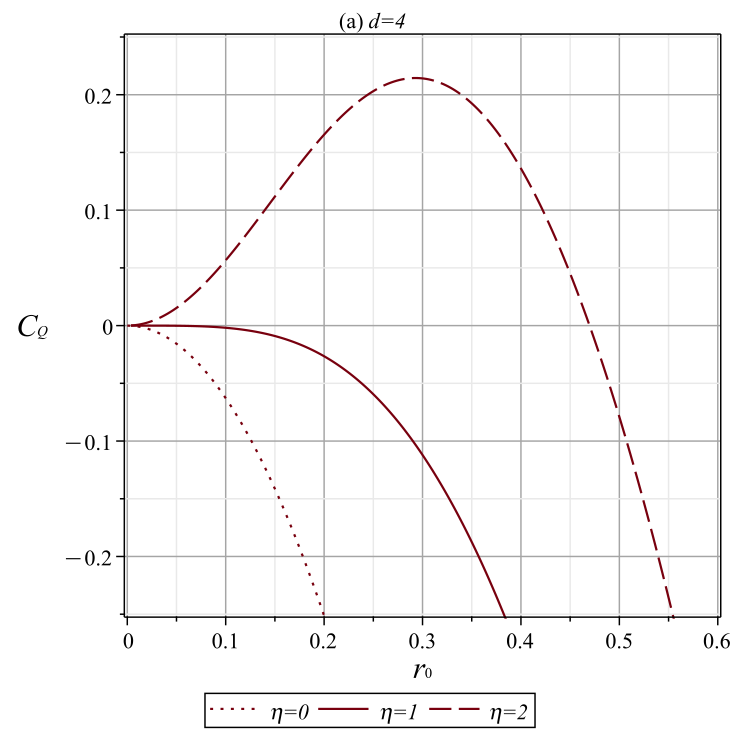}}
  \resizebox{0.45\textwidth}{!}{%
  \includegraphics{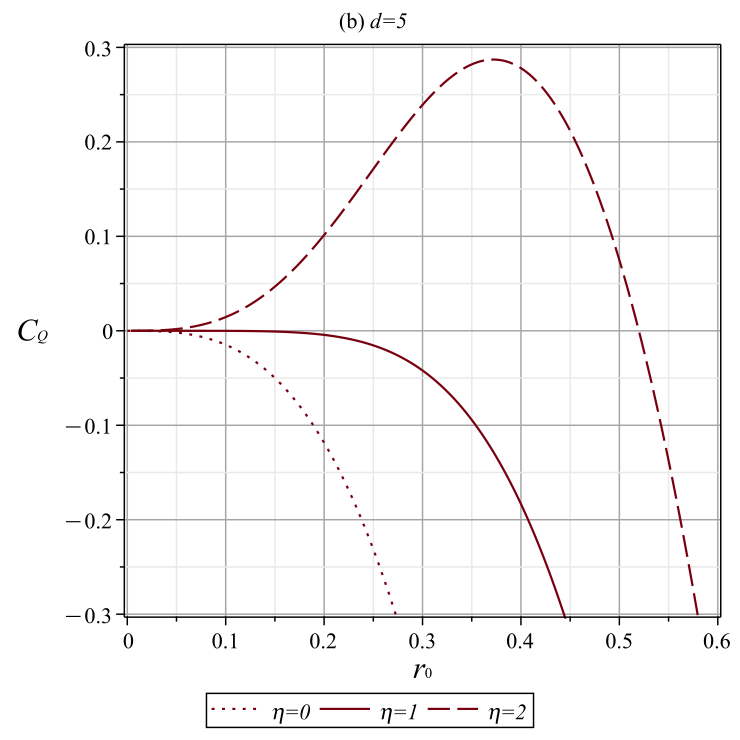}}
\vspace{0.5cm}       
\caption{Specific heat versus $r_{0}$ with $G_d=1$ for Schwarzschild black hole in 4 and 5 dimensions.}
\label{fig:C}       
\end{figure}

In plots of Fig. \ref{fig:C}, we can see the behavior of the specific heat in terms of $r_{0}$ and see that in the presence of the exponential correction, the Schwarzschild black hole may be stable at small radii. Although, we have only presented two cases, i.e.,  $d=4$ (Fig. \ref{fig:C} (a)) and $d=5$ (Fig. \ref{fig:C} (b)), we find similar behavior in other dimensions.  From the Fig. \ref{fig:C} we can see a second order phase transition may happen to have a stable black hole at small horizon radius. However, this stable phase only exists for $\eta>1$, which is not our case of interest. Therefore, we find that small $d$-dimensional Schwarzschild black holes are also instable for $0<\eta<1$, and this hold for the four dimensional Schwarzschild black hole too.

\section{Conclusion}\label{sec5}
 In this paper, we have analyzed the effects of quantum gravitational corrections on  higher-dimensional Schwarzschild black holes. This was achieved by calculating the Kullback-Leibler divergence, which measures the difference between the original probability distribution of particles emitted during black hole evaporation and the corrected probability distribution influenced by quantum gravitational effects. Our results demonstrate  that the Kullback-Leibler divergence  depends on both the dimensionality and the scale of the black holes. Initially, higher-dimensional black holes exhibit a larger divergence compared to lower-dimensional ones, indicating that quantum gravitational corrections are more pronounced in higher dimensions. However, as the black hole approaches a critical scale near the Planck regime, this behavior reverses. At this stage, quantum fluctuations dominate, and new quantum gravitational degrees of freedom become significant, leading to a greater divergence for lower-dimensional black holes.

This behavior is further illuminated through the analysis of Fisher information, which quantifies the sensitivity of an observable probability distribution to changes in a parameter of interest. This parameter here quantifies the strength of quantum gravitational corrections. We observe that Fisher information about quantum gravitational effects follows a similar pattern: it is initially larger for higher-dimensional black holes but becomes more significant for lower-dimensional black holes near the Planck scale. Moreover, as the black hole evaporates, the Fisher information attains a finite maximum value, providing an upper bound on the retrievable information about quantum gravitational corrections. This is an information theoretical result, which is equivalent to the heuristic breaking of spacetime around Planck scale. Such a breaking would also produce a bound on the Fisher information about quantum gravitational effects. Here, we have obtained such a bound in a rigorous way, using a black-box approach. This is done by analyzing the probability distribution of particles emitted during the evaporation of a black hole.  
Furthermore, the maximum value of Fisher information and the corresponding peak in the Kullback-Leibler divergence highlight the scale at which quantum gravitational corrections are most detectable. This occurs at the critical point where quantum fluctuations  dominate. 

This work demonstrates fundamental limitations on the information retrievable about both the black hole's initial state and the quantum gravitational corrections that govern its evolution. The finite maximum value of Fisher information, inversely proportional to the strength of quantum gravitational effects, quantifies the constraints imposed by quantum gravity. This boundedness addresses the black hole information paradox by suggesting that quantum gravitational corrections could encode a finite information. However, it simultaneously introduces a meta-information limit on information, wherein the ability to extract information about the nature of quantum gravity itself is intrinsically limited.
This self-concealing property of quantum gravitational corrections implies that they predict an ultraviolet limit on the information  near the Planck scale.   They obscure details about the underlying quantum gravitational degrees of freedom. This  behavior creates an intrinsic epistemic boundary within quantum gravity, reflecting its self-referential nature and presenting challenges for reconstructing a complete and consistent  quantum theory of gravity. This resembles inconsistencies theorems in formal logic and could in principle be connect to them.    

We have also shown that quantum gravitational corrections, encapsulated by the parameter \(\eta\), can significantly alter the thermodynamic stability of Schwarzschild black holes, particularly at small horizon radii where quantum effects dominate. These corrections modify key thermodynamic quantities such as Helmholtz free energy, internal energy, and specific heat, revealing the possibility of stable black hole configurations for \(\eta \geq e^{S_d}\). However, in the regime \(0 < \eta < 1\), small black holes remain thermodynamically unstable, and only for \(\eta > 1\) can stable phases with a second-order phase transition emerge, though this is outside our primary focus. Additionally, a trade-off between stability and sensitivity to quantum corrections is highlighted through Fisher Information, suggesting that as quantum effects stabilize the black hole, the system becomes less sensitive to perturbations. This work has the potential of using quantum gravitationally corrections to  enhance our understanding of black hole physics, and even computational limitations of quantum gravity.

This work highlights several open problems and future directions that could extend the understanding of quantum gravitational corrections and their implications for black hole physics. One interesting possibility  is the analysis of quantum gravitational effects in black holes with additional parameters, such as rotating Kerr black holes or charged Reissner-Nordström black holes. These cases introduce angular momentum and electric charge, which could interact with quantum gravitational corrections in interesting ways. This could potentially modifying the emission of particles,  and dynamics of evaporation. Furthermore, black holes in asymptotically anti-de Sitter (AdS) or de Sitter (dS) spacetimes provide opportunities to explore how quantum corrections using  holographic dualities or in cosmological settings, respectively. For AdS spacetimes, such corrections could have  implications for the AdS/CFT correspondence, particularly near the Planck scale, where deviations from classical physics become significant. In dS spacetimes, the interplay between black hole evaporation and cosmological expansion raises new questions about how quantum gravitational effects influence horizons in an accelerating universe.

Another important extension involves black holes in modified theories of gravity, such as those incorporating higher curvature corrections (e.g., Einstein-Gauss-Bonnet gravity) or alternative metric formulations. These scenarios could provide insights into the behavior of quantum gravitational corrections in regimes beyond standard general relativity. Similarly, investigating time-dependent black holes and dynamical horizons, such as those encountered in black hole mergers or other highly non-equilibrium processes, could uncover the influence of quantum corrections in dynamically evolving systems. Applications to primordial black holes are particularly interesting, as these small black holes are expected to approach the Planck scale during evaporation, making them ideal candidates for studying the observational consequences of quantum gravitational effects. Such studies could connect to broader astrophysical phenomena, including the potential role of primordial black hole remnants in dark matter models or the detection of modified Hawking radiation spectra.

Observationally, the  Fisher information and Kullback-Leibler divergence  may offer critical insights into detecting quantum gravitational effects. These values represent the optimal regime for experimental sensitivity to quantum corrections. Beyond specific black hole types and observational considerations, this work opens new avenues for exploring quantum information theoretical perspectives on black hole physics. Quantities such as quantum complexity, entanglement entropy, and mutual information may complement Fisher information in characterizing the influence of quantum gravity on black hole systems. These approaches could also provide a deeper understanding of the information paradox and its resolution, particularly in terms of how quantum gravity encodes and conceals information.
The implications of this study are not limited to black hole systems. They raise broader questions about the fundamental nature of spacetime and the self-referential behavior of quantum gravity, where corrections can obscuring the theoretical framework underlying them.

\section*{Appendix: Jacobson Formalism}
   {In order to explicit construct the quantum geometry, we will now use the   Jacobson formalism \cite{gr12, jacob}. This can be done by 
considering a two-surface element $\cal P$ at point $p$ with an associated Killing field $\chi^a$ generating orthogonal boosts. The system exhibits an Unruh temperature given by $\hbar \kappa/2\pi$, where $\kappa$, which represents the acceleration along the Killing orbit. Heat flow is characterized by the boost-energy current $T_{ab}\chi^a$, with a local Rindler horizon through $p$ generated by $\chi^a$. The heat flux $\cal H$ through $\cal P$ can be expressed as,
\begin{equation}
\delta Q = \int_{\cal H} T_{ab} \chi^a d\Sigma^b,
\label{eq:heat_flux}
\end{equation}
where $d\Sigma^a = k^a d\lambda d\cal A$, with $k^a$ denoting the horizon's tangent vector. The affine parameter $\lambda$ vanishes at $\cal P$ and takes negative values in its past. Using the area element $d\cal A$, we obtain,
\begin{equation}
\delta Q = -\kappa \int_{\cal H} \lambda T_{ab} k^a k^b d\lambda d\cal A.
\label{eq:heat_area}
\end{equation}
The horizon's expansion $\theta$ can be related  to the area's variation as
\begin{equation}
\delta {\cal A} = -\kappa \int_{\cal H} \theta d\lambda d\cal A.
\label{eq:area_variation}
\end{equation}
The Raychaudhuri equation which governs the horizon's evolution is $
{d\theta}/{d\lambda} = -\theta^2/2 - \sigma^2 - R_{ab}k^ak^b.
$  
Now integration, we obtain the following expression 
\begin{equation}
\delta {\cal A} = -\int_{\cal H} \lambda R_{ab} k^a k^b d\lambda d\cal A
\label{eq:integrated_area}
\end{equation}
The thermodynamic relation $\delta Q = TdS = (\hbar\kappa/2\pi)\xi\delta\cal A$ holds when $T_{ab}k^ak^b = (\hbar\xi/2\pi) R_{ab}k^ak^b$ for all null vectors $k^a$. This leads to,
\begin{equation}
R_{ab} - \frac{1}{2}R g_{ab} + \Lambda g_{ab} = \frac{2\pi}{\hbar\xi}T_{ab},
\label{eq:einstein}
\end{equation}
where $\xi$ relates to Newton's constant as $G_{d} = \frac{1}{4\hbar\xi}$. In the paper, we applied this formalism to corrected entropy by defining a corrected area, where the entropy obtained from the corrected area coincides with the quantum gravitationally corrected entropy.    }

\section {Acknowledgements}
 \.{I}.S. would like to acknowledge networking support of COST Actions CA18108, CA21106,CA23130, and CA22113. He also thanks T\"{U}B\.{I}TAK and SCOAP3 for their support.

\section*{Data Availability}
There is no associated data in this paper.


\begin{thebibliography}{200}

\bibitem{le1}A.~F.~Ali, S.~Das and E.~C.~Vagenas,
Phys. Rev. D {84}, 044013 (2011)
\bibitem{le2}M.~F.~Wondrak and M.~Bleicher, Symmetry  {11}, no.12, 1478 (2019)
\bibitem{le4}O.~Bertolami, J.~G.~Rosa, C.~M.~L.~de Aragao, P.~Castorina and D.~Zappala,
Phys. Rev. D  {72}, 025010 (2005)
\bibitem{le5}M.~Khodadi, K.~Nozari, S.~Dey, A.~Bhat and M.~Faizal,
Sci. Rep. {8}, no.1, 1659 (2018)
\bibitem{le6}I.~Pikovski, M.~R.~Vanner, M.~Aspelmeyer, M.~Kim and C.~Brukner,
Nature Phys.  {8}, 393-397 (2012)
\bibitem{le7}S.~Das and E.~C.~Vagenas,
Phys. Rev. Lett.  {101}, 221301 (2008)
\bibitem{qge12}G.~Amelino-Camelia, J.~R.~Ellis, N.~E.~Mavromatos, D.~V.~Nanopoulos and S.~Sarkar,
Nature  {393}, 763-765 (1998)
\bibitem{qge14}A.~Belenchia, D.~M.~T.~Benincasa, S.~Liberati, F.~Marin, F.~Marino and A.~Ortolan,
Phys. Rev. Lett. {116}, no.16, 161303 (2016)
\bibitem{qge15} S.~E.~Perkins, N.~Yunes and E.~Berti,
Phys. Rev. D {103}, no.4, 044024 (2021)
\bibitem{qge16}R.~B.~Mann, S.~Murk and D.~R.~Terno,
Int. J. Mod. Phys. D  {31}, no.09, 2230015 (2022)
\bibitem{pw} M. K. Parikh and F. Wilczek, Phys. Rev. Lett. 85, 5042 (2000)
 \bibitem{pw1}P. Ghosal and R. Ray, Phys. Rev. D 105, 124016 (2022)
\bibitem{kl1}I. Csiszar, Ann. Probab. 3,  146 (1975)
\bibitem{kl2} S.  Kullback,  Ame. Statistician 41,  340 (1987) 
\bibitem{fish1}B.~R.~Frieden and B.~H.~Soffer,
Phys. Rev. E  {52}, 2274 (1995)
\bibitem{fish2}T.~Cokelaer, Class. Quant. Grav.  {25}, 184007 (2008)
\bibitem{k1} L.~Randall and R.~Sundrum,
 Phys. Rev. Lett.  {83}, 4690-4693 (1999)

\bibitem{k2}A.~Chamblin and G.~C.~Nayak,
Phys. Rev. D  {66}, 091901 (2002)
\bibitem{b1}
B.~Pourhassan and M.~Faizal, JHEP {10}, 050 (2021)
\bibitem{b2}
B.~Pourhassan, S.~Dey, S.~Chougule and M.~Faizal, Class. Quant. Grav. {37}, 135004 (2020)
\bibitem{ed1} J.~Ponce de Leon and N.~Cruz, Gen. Rel. Grav.  {32}, 1207-1216 (2000)
\bibitem{ed2}A.~Ishibashi and H.~Kodama, Prog. Theor. Phys.  {110}, 901-919 (2003)
\bibitem{ed3}E.~Jung and D.~K.~Park, Nucl. Phys. B  {766}, 269-283 (2007)
\bibitem{ed4}P.~Kanti, T.~Pappas and N.~Pappas, Phys. Rev. D  {90}, no.12, 124077 (2014)
\bibitem{j6}
B.~Pourhassan, H.~Aounallah, M.~Faizal, S.~Upadhyay, S.~Soroushfar, Y.~O.~Aitenov and S.~S.~Wani, JHEP  {05}, 030 (2022) 
\bibitem{qge17}B.~Koch, M.~Bleicher and S.~Hossenfelder,
 JHEP  {10}, 053 (2005)
\bibitem{qge18}A.~Chamblin and G.~C.~Nayak,
Phys. Rev. D  {66}, 091901 (2002)
\bibitem{qge19}A.~F.~Ali, M.~Faizal and M.~M.~Khalil,
Phys. Lett. B  {743}, 295-300 (2015)
\bibitem{qge10}B.~Pourhassan, S.~S.~Wani and M.~Faizal,
Nucl. Phys. B  {960}, 115190 (2020)
\bibitem{1a00}
J.D. Bekenstein, Phys. Rev. D 9, 3292 (1974)
\bibitem{1b}
J.D. Bekenstein, Phys. Rev. D 7, 2333 (1973)
\bibitem{1c}
S.W. Hawking, Nature (London) 248, 30 (1974)
\bibitem{1a0}
S.W. Hawking, Commun. Math. Phys. 43, 199 (1975)
\bibitem{Kanti:2014dxa}
P.~Kanti, T.~Pappas and N.~Pappas,
Phys. Rev. D  {90}, no.12, 124077 (2014)
\bibitem{Pappas:2016ovo}
T.~Pappas, P.~Kanti and N.~Pappas,
Phys. Rev. D  {94}, no.2, 024035 (2016)
\bibitem{Sakalli:2022xrb}
\.I.~Sakalli and S.~Kanzi, 
Turk. J. Phys.  {46}, no.2, 51-103 (2022)
\bibitem{32}
S. Das, P. Majumdar and R. K. Bhaduri,   Class. Quantum Grav. 19, 2355 (2002)
\bibitem{32a}
S. Upadhyay, B. Pourhassan and H. Farahani, Phys. Rev. D 95, 106014 (2017)
\bibitem{32b}
A.~Jawad, Class. Quant. Grav.  {37}, 185020 (2020)
\bibitem{32c}
M.~Rostami, J.~Sadeghi, S.~Miraboutalebi and B.~Pourhassan,
Annals Phys. {429}, 168488 (2021)
  \bibitem{32d}
J.~Sadeghi, B.~Pourhassan and M.~Rostami, Phys. Rev. D  {94}, 064006 (2016)
\bibitem{gr12}
T. Jacobson, Phys. Rev. Lett. 75, 1260 (1995)
\bibitem{gr14}
M.~Faizal, A.~Ashour, M.~Alcheikh, L.~Alasfar, S.~Alsaleh and A.~Mahroussah, Eur. Phys. J. C {77}, 608 (2017)
\bibitem{12P}
B. Pourhassan and M. Faizal,   Europhys. Lett.\, {111}, 40006 (2015)
\bibitem{12ap}
J. Sadeghi, B. Pourhassan and F. Rahimi,  Can. J. Phys.\, {92}, 1638 (2014)
\bibitem{13p}
M. Faizal and B. Pourhassan,  Phys. Lett. B {751}, 487 (2015)
\bibitem{19p}
B. Pourhassan and M. Faizal,  Nucl. Phys. B {913}, 834 (2016)
\bibitem{4p}
B.~Pourhassan, M.~Faizal, S.~Upadhyay and L.~A.~Asfar,
Eur. Phys. J. C  {77}, no.8, 555 (2017)
\bibitem{hbh} N. A. Shah, A. Naqash,  A. K. Khan, R. F. Shah, and S. Lone, JHAP 3 (2), 17-30 (2023)
\bibitem{4a}
L. Susskind, J. Math. Phys. 36, 6377  (1995)
\bibitem{5a}
R. Bousso, Rev. Mod. Phys. 74, 825 (2002)
\bibitem{6a}
D. Bak and S. J. Rey, Class. Quant. Grav. 17, L1  (2000)
\bibitem{7a}
S. K. Rama, Phys. Lett. B 457, 268  (1999)
\bibitem{18}
S. Hemming and L. Thorlacius,   JHEP 11, 086 (2007)
\bibitem{18a}
R.~Gregory, S.~F.~Ross and R.~Zegers, JHEP  {09}, 029  (2008)
\bibitem{18b}
J.~V.~Rocha, JHEP {08}, 075  (2008)
\bibitem{18c}
Z.~H.~Li, B.~Hu and R.~G.~Cai, Phys. Rev. D {77}, 104032 (2008)
\bibitem{18d}
K.~Saraswat and N.~Afshordi, JHEP  136, {04} (2020)
\bibitem{wald1}R. M. Wald, Phys. Rev. D 48, 3427 (1993)
\bibitem{wald2} A. Sen,  JHEP 0509, 038 (2005)
\bibitem{6ab}
G. Lifschytz and M. E. Ortiz, Nucl.Phys. B 486, 131 (1997)
\bibitem{7ab}
S.~Mahapatra, Eur. Phys. J. C  {78}, 23 (2018)
\bibitem{7ba}
C.~Keeler, F.~Larsen and P.~Lisbao, Phys. Rev. D   043011, 90 (2014)
\bibitem{Dabholkar}
A. Dabholkar, J. Gomes and S. Murthy,   JHEP 03, 074 (2015)
\bibitem{ds12}
A.~Dabholkar, J.~Gomes and S.~Murthy, JHEP 04, 062  (2013)
\bibitem{ds14}
S. Murthy and B. Pioline, JHEP 09, 022 (2009)
\bibitem{main}
A. Chatterjee and A. Ghosh, Phys. Rev. Lett. {125}, 041302 (2020) 
\bibitem{bia}B.~Pourhassan, J. Stat. Mech. {2107}, 073102 (2021)
\bibitem{bia1}B.~Pourhassan, M.~Dehghani, M.~Faizal and S.~Dey,
Class. Quant. Grav.  {38}, no.10, 105001 (2021)
\bibitem{j1}
B.~Pourhassan, M.~Atashi, H.~Aounallah, S.~S.~Wani, M.~Faizal and B.~Majumder, Nucl. Phys. B  {980}, 115842 (2022)
\bibitem{j2}
B.~Pourhassan, S.~S.~Wani, S.~Soroushfar and M.~Faizal, JHEP  {10}, 027 (2021)

{ \bibitem{Mathur:2009hf}
S.~D.~Mathur,
 Class. Quant. Grav.  {26}, 224001 (2009)

\bibitem{Almheiri:2019psf}
A.~Almheiri, N.~Engelhardt, D.~Marolf and H.~Maxfield,
 JHEP  {12}, 063 (2019)

\bibitem{Harlow:2014yka}
D.~Harlow,
 Rev. Mod. Phys.  {88}, 015002 (2016)

\bibitem{Preskill:1992tc}
J.~Preskill,
 [arXiv:hep-th/9209058 [hep-th]]

\bibitem{Sakalli:2012zy}
I.~Sakalli, M.~Halilsoy and H.~Pasaoglu,
 Astrophys. Space Sci.  {340}, no.1, 155-160 (2012)

\bibitem{Sakalli:2010yy}
I.~Sakalli, M.~Halilsoy and H.~Pasaoglu,
Int. J. Theor. Phys.  {50}, 3212-3224 (2011)}
\bibitem{information}P.~Chen, Y.~C.~Ong and D.~h.~Yeom,
Phys. Rept.  {603}, 1-45 (2015) 
\bibitem{information12}S.~L.~Braunstein and A.~K.~Pati,
Phys. Rev. Lett.  {98}, 080502 (2007)
 
 
 \bibitem{qg1}A.~F.~Ali and M.~M.~Khalil,
Nucl. Phys. B {909}, 173-185 (2016)
\bibitem{qg2}E.~C.~Vagenas, L.~Alasfar, S.~M.~Alsaleh and A.~F.~Ali,
Nucl. Phys. B  {931}, 72-78 (2018)
\bibitem{qg3}J.~Magueijo and L.~Smolin,
Class. Quant. Grav.  {21}, 1725-1736 (2004)
 \bibitem{qg4}A.~F.~Ali,
Phys. Rev. D {89}, no.10, 104040 (2014)
\bibitem{qg5}S.~H.~Mehdipour,
Phys. Rev. D {81}, 124049 (2010)
\bibitem{qg6} P.~Nicolini, E.~Spallucci and M.~F.~Wondrak, Phys. Lett. B  {797}, 134888 (2019)
\bibitem{SPR}
R.~C.~Myers and M.~J.~Perry,
Annals Phys.  {172}, 304 (1986)
\bibitem{min1}
S.~Doplicher, K.~Fredenhagen and J.~E.~Roberts,
Commun. Math. Phys. {172}, 187-220 (1995)
 \bibitem{min2}T. Padmanabhan, Phys. Rev. Lett. 78, 1854 (1997)
\bibitem{min3}M.~Faizal,
Int. J. Mod. Phys. A {38}, no.35n36, 2350188 (2023)
\bibitem{min4}A.~F.~Ali,
Class. Quant. Grav.  28, 065013 (2011)
\bibitem{mi}A.~Kempf, G.~Mangano and R.~B.~Mann,
Phys. Rev. D {52}, 1108-1118 (1995)
\bibitem{Das:2010sj}
S.~Das and E.~C.~Vagenas,
Phys. Rev. Lett. {104}, 119002 (2010)
\bibitem{h1}M.~Maggiore,
Phys. Lett. B  {304}, 65-69 (1993)
\bibitem{h2}L.~J.~Garay,
Int. J. Mod. Phys. A  {10}, 145-166 (1995)
\bibitem{Pikovski:2011zk}
I.~Pikovski, M.~R.~Vanner, M.~Aspelmeyer, M.~Kim and C.~Brukner,
Nature Phys.  {8}, 393-397 (2012)
\bibitem{hn}F.~Marin, F.~Marino, M.~Bonaldi, M.~Cerdonio, L.~Conti, P.~Falferi, R.~Mezzena, A.~Ortolan, G.~A.~Prodi and L.~Taffarello, \textit{et al.}
Nature Phys.  {9}, 71-73 (2013)
 \bibitem{Das:2011tq}
S.~Das and R.~B.~Mann,
Phys. Lett. B {704}, 596-599 (2011)
\bibitem{Bosso:2016ycv}
P.~Bosso, S.~Das, I.~Pikovski and M.~R.~Vanner,
Phys. Rev. A {96}, no.2, 023849 (2017)
\bibitem{Ali:2011fa}
A.~F.~Ali, S.~Das and E.~C.~Vagenas,
Phys. Rev. D  {84}, 044013 (2011)
\bibitem{Pedram:2011xj}
P.~Pedram, K.~Nozari and S.~H.~Taheri,
JHEP  {03}, 093 (2011)
\bibitem{Bosso:2023aht}
P.~Bosso, G.~G.~Luciano, L.~Petruzziello and F.~Wagner,
Class. Quant. Grav.  {40}, no.19, 195014 (2023)
\bibitem{1}A.~J.~M.~Medved and E.~C.~Vagenas,
Phys. Rev. D  {70}, 124021 (2004)
\bibitem{2}P.~Chen, Y.~C.~Ong and D.~h.~Yeom,
Phys. Rept.  {603}, 1-45 (2015)
\bibitem{3} A.~Alonso-Serrano, M.~P.~Dabrowski and H.~Gohar, Phys. Rev. D  {97}, no.4, 044029 (2018)
\bibitem{4}D.~J.~Gogoi and U.~D.~Goswami,
JCAP {06}, no.06, 029 (2022)
\bibitem{5}Y.~C.~Ong,
JHEP {10}, 195 (2018)
\bibitem{6}W.~Kim and J.~J.~Oh, JHEP {01}, 034 (2008)
\bibitem{Nozari:2009nr} K.~Nozari and S.~H.~Mehdipour,
JHEP  {03}, 061 (2009)
\bibitem{Nozari:2012nf} K.~Nozari and S.~Saghafi,
JHEP {11}, 005 (2012)
\bibitem{Chen:2013ssa}
D.~Y.~Chen, Q.~Q.~Jiang, P.~Wang and H.~Yang,
JHEP {11}, 176 (2013)
\bibitem{Arzano:2005rs}
M.~Arzano, A.~J.~M.~Medved and E.~C.~Vagenas,
JHEP  {09}, 037 (2005)
\bibitem{Carr:2015nqa}
B.~J.~Carr, J.~Mureika and P.~Nicolini,
JHEP {07}, 052 (2015) 
\bibitem{Ling:2021olm}
Y.~Ling and M.~H.~Wu,
 Class. Quant. Grav.  {40}, no.7, 075009 (2023)
\bibitem{SPR2}F. R. Tangherlini,  Nuovo Cim. 27, 636 (1963)
\bibitem{Argyres}
P.~C.~Argyres, S.~Dimopoulos and J.~March-Russell,
Phys. Lett. B {441}, 96-104 (1998)
\bibitem{exp12}S.~Soroushfar, H.~Farahani and S.~Upadhyay, Phys. Dark Univ.  {42}, 101272 (2023)
\bibitem{exp14}A.~Jawad and U.~Zafar, Nucl. Phys. B  {992}, 116231 (2023)
\bibitem{hi12} 
B.~Pourhassan, K.~Kokabi and Z.~Sabery,
Annals Phys. {399}, 181-192 (2018)
\bibitem{hi14}B.~Pourhassan,
Eur. Phys. J. C {79},  740 (2019)
\bibitem{fish4}E. Makalic and  D. F.  Schmidt,    IEEE Signal Proc. Lett. 17, 391-393 (2009)
\bibitem{fish5} S. I. Costa, S. A. Santos, and J. E. Strapasson, Discrete App. Math. 197, 59-69 (2015)
\bibitem{jacob}
T. Jacobson, A. Mohd, Phys. Rev. D 92, 124010 (2015)


\bibitem{in12}M.~Faizal, A.~Shabir and A.~K.~Khan,
Int. J. Theor. Phys.  {63}, no.11, 290 (2024)
\bibitem{in14}M.~Faizal, A.~Shabir and A.~K.~Khan,
EPL  {148}, no.3, 39001 (2024) 
\bibitem{int16}M.~Faizal, A.~Shabir and A.~K.~Khan,
Nucl. Phys. B {1010}, 116774 (2025)
\end{thebibliography}
\end{document}